\documentclass[runningheads]{llncs}

\usepackage{amssymb}
\usepackage{graphicx}

\usepackage{url}
\urldef{\mailsa}\path|{jmc,|
\urldef{\mailsb}\path|eric.mahe}@massiverand.com|    
\newcommand{\keywords}[1]{\par\addvspace\baselineskip
\noindent\keywordname\enspace\ignorespaces#1}
\usepackage{makeidx}  
\usepackage[latin1]{inputenc}
\usepackage[T1]{fontenc}
\begin{document}
\mainmatter  
\title{Secrets from the GPU}
\titlerunning{Secrets from the GPU}  
%
\author{Jean-Marie Chauvet \and Eric Mahé}
\authorrunning{J.-M. Chauvet et al.} 
%
\tocauthor{Jean-Marie Chauvet, Eric Mahé}
\institute{MassiveRand, Inc.\\
\mailsa\mailsb \\
\texttt{http://www.massiverand.com} \\
62, ave. Pierre Grenier, 92100 Boulogne-Billancourt, France}
\maketitle              
\begin{abstract}
Acceleration of cryptographic applications on massively parallel computing platforms, such as Graphics Processing Units (GPUs), becomes a real challenge as their decreasing cost and mass production makes practical implementations attractive.  We propose a layered trusted architecture integrating random bits generation and parallelized RSA cryptographic computations on such platforms. The GPU-resident, three-tier, MR architecture consists of a RBG, using the GPU as a deep entropy pool; a bignum  modular arithmetic library using the Residue Number System; and GPU APIs for RSA key generation, encryption and decryption.  Evaluation results of an experimental OpenCL implementation show a 32-40 GB/s throughput of random integers, and encryptions with up to 16,128-bit long exponents on a commercial mid-range GPUs. This suggests an ubiquitous solution for autonomous trusted architectures combining low cost and high throughput. \dots
\keywords{Cryptography, Residue Number System, GPU, Random Bit Generator, RSA}
\end{abstract}
\section{Introduction}
\label{sec:introduction}
 Recent systemic studies of trust with asymmetric cryptography usage led to questioning the assumption that sufficient randomnessis available each time public keys are generated \cite{Lenstra2012}. The issue of the \emph{quality} of random bits used in key generation and the depth of their associated entropy pools is of course thre prominent one raised in these studies. The extent to which public keys are in fact shared among unrelated parties also points to the need for fast, low-cost, readily available, autonomous key generation and encryption-decryption capability.

In this contribution, we focus on the efficient realization of the computationally expensive operations in asymmetric cryptosystems on off-the-shelf GPUs. More precisely, we present improved and novel implementations employing GPUs both as entropy pool and accelerator for RSA and DSA cryptosystems in a complete three-tiered architecture. The MassiveRand (MR) Architecture consists of three functional layers: (i) MR-TRNG, a non-deterministic random bits generator (TRBG); (ii) MR-MOD, a parallelized efficient implementation of modular arithmetic on big integers; and (iii) MR-RSA, an application layer offering a simple API for RSA key generation and message encryption/decryption.

The novelty of the MR Architecture resides in the dual use of mass-marketed, low-cost hardware, namely GPUs -- as now found in a large range of computing devices ranging from high-end servers to smartphones and tablets -- as a deep entropy pool for an innovative TRBG tightly integrated with an on-board, efficiently parallelized cryptographic library for a variety of applications. The architecture delivers high-bandwith, high-quality secrets on a broad set of computing devices.

\section{GPU Hardware as an Ubiquitous, Deep, Entropy Pool}
\label{sec:EntropyPool}
 A typical GPU architecture \cite{OwensGPU2008} consists of several general purpose scalar processors grouped in multiprocessor cores with a hierarchy of global, local and resident (cache) memory allowing several levels of data parallelism and optimization of parallel computations. It is becoming increasing common to use a GPU as a modified form of stream processor. This idea turns the massive computational power of a modern graphics accelerator's shader pipeline into general-purpose computing power, as opposed to being hard-wired solely to do graphical operations.

On the other hand, low cost, mass production of GPUs make them ubiquitous, not only on PCs, but on widely released devices such as laptops, smartphones and tablets as well as in special architectures for high-performance computing, with an increasing performance-price ratio every year.

The first layer (MR-TRNG) of the cryptography architecture proposed in this paper leverages the GPU as a specific random bit generator (RBG) hardware. In accordance with standards, e.g. FIPS 140-2 \cite{FIPS:2001:SRC,SP80090C}, the GPU hardware is used here as a \emph{non-deterministic} RBG, producing an output that is dependent on an unpredictable (hardware) source that is ``outside human control''. The unpredictable stream of bits is seeded as an unguessable input key to an approved deterministic RBG (DRBG or pseudo-RNG). The collection of entropy and the DRBG are both GPU-bound programs, known as \emph{kernels}.

In contrast to the current view that the task of a DRBG is simply to distill out sufficient entropy for all outputs and queue it up for use--shutting down completely or generating predictable pseudo-random output when it does not see enough entropy--MR-TRNG innovative use of GPU hardware ensures that sufficient entropy is continuously collected for the production of unguessable keys.
\section{Cryptographic GPU Computing}
\label{sec:Gryptography}
 For trust, RSA cryptographic computations use large key sizes \cite{FIPS1863}, usually 1024-, 2048- and 3072-bit long integers, so called \emph{bignums}. In addition to the difficulty of computing modular arithmetic operations on bignums, the algorithms traditionally used in RSA cryptography \cite{HAC} do not easily translate to efficient implementation on highly parallel architectures. Multiplication and exponentiation by bignums, in particular, are the most demanding. These basic operations are conducted stepwise, in several bignum parsing phases, inducing complex and heavy data dependencies between steps. In order to overcome these data dependency obstacles, the proposed architecture relies on the Residue Number System (RNS) representation of bignums which reduces bignum operations to easily parallelized small modulus (32-bit or 64-bit) arithmetic computations.
\subsection{Modular Arithmetic Operations on GPU}
 The main principle of the RNS is to consider several \emph{moduli} $m_1$, $m_2$, ..., $m_n$ that contain no common factors--they are \emph{coprimes} to each other--and to work indirectly with residues of bignums modulo each $m_i$ rather than directly with the bignum itself. Hence, in the RNS, a bignum $X$ is represented by $(x_1, x_2, ..., x_n)$, the list of its residues $x_i=X\:mod\:m_i$.

The fact that this representation is unique, provided $X < M = \prod m_i$, is the Chinese Remainders Theorem \cite{TheArt}. In this convenient representation addition, subtraction and multiplication on bignums can be performed in parallel modulo each $m_i$. These independent streams of computation, or \emph{channels}, are directly mapped onto \emph{threads} on the GPU computing hardware.

Conversions from binary to the RNS are executed by computing the residues $x_i$ on each channel which requires only small integer operations. The opposite conversion, from the RNS to binary, however, is more involved. Two major methods are available: translating first to an intermediary Mixed Radix System (MRS), which although sequential in nature may be computed separately on each channel \cite{TheArt}; or using the Chinese Remainders Theorem which lends itself to proper parallelization provided an extra modulus for intermediate calculations.

RSA arithmetic requires furthermore modular operations, namely multiplication and exponentiation, to be performed efficiently. MR-MOD uses Montgomery modular multiplication \cite{Bajard2001}. This method simultaneously performs multiplication and reduction by a (bignum) modulus $N$, in a so-called Montgomery domain characterized by a large integer $R$. Bignum operands $X$, $Y$ are replaced by $\tilde{X}=XR$, $\tilde{Y}=YR$ and the Montgomery modular multiplication computes $\tilde{Z}=\tilde{X}\tilde{Y}R^{-1}\:mod\:N$, provided again that $N<R$. By choosing $R=M=\prod m_i$, single steps are performed independently on each RNS channel, consuming RNS representations of $X$, $Y$, and $N$. Because translating bignums to and from the Montgomery domain is costly, MR-MOD performs as many operations as possible in the Montgomery domain. Such is the case for exponentiation, implemented with traditional square-and-multiply algorithms \cite{HAC},  efficiently chaining Montgomery modular multiplications. In addition, pre-computations help moving many multiplications outside the main exponentiation loop \cite{Bajard2010} thus improving overall performance.

Modular inversion is required to compute the private exponent $d$ of the RSA key. Inversion is also used in the constructive prime generation methods used in the upper RSA layer of the proposed architecture. The current implementation implements the Arazi inversion formula which given $e$ and $f$ coprime positive integers yields $d=e^{-1}\:mod\:f$ as $d=\frac{1+f(-f^{-1}\:mod\:e)}{e}$. The simplification in this formula relies on $e$ being a small integer.  The modular inverse of bignum $f$ is computed by first converting $f$ to the RNS and then inverting by the Arazi formula, in parallel on each channel. A careful choice of coprime bases for the RNS leads to additional performance increases \cite{Bajard2009}.

Once the RNS bases are chosen, both base extension and intermediate calculations in the Montgomery modular multiplication use values which are constant throughout RNS operations. In order to expedite GPU computations, these constants are pre-computed and installed permanently in GPU memory at initialization time. In the current implementation, each base counts 128 32-bit prime integers, sufficient for RSA keys up to 3,968-bit long, slightly over the 3072-bit maximum length in the NIST recommendation \cite{SP80090C}. (On modern GPUs, blocks may run up to 256 threads so that doubling the base size and switching to 64-bit coprimes would give us capability for up to 16,128-bit long RSA keys, without much computation overhead.)
\subsection{Primality}
 When efficiency is not a primary concern, the usual way to generate a random prime number is to select a random number $p$ and test it for primality. This, in essence, describes the operations of the lower-tier of the third and last layer of the proposed architecture, MR-RSA. Random numbers issued by GPU-based MR-TRNG are tested for primality using the MR-MOD modular arithmetic kernels, both generation and computations being executed in parallel on the GPU.  Prime generation algorithms rely on primality, or compositeness, testing \cite{Schoof2008}. In order to keep generated prime numbers on board the GPU in the trust management context, without transferring any other integer than the public RSA key and exponent to the host CPU, primality testing in the Montgomery domain has been implemented as a dedicated GPU kernel. In accordance with standards and recommendations \cite{FIPS1863}, the kernel executes a Miller-Rabin test with a user-parameterized number of iterations.
\subsection{RSA Key Generation, Encryption and Decryption}
Finally the upper-tier of the MR-RSA layer simply provides a basic RSA API for use in cryptographic applications. Note again that the GPU acts both as the entropy pool for random bits generation and specific computing hardware for key generation, encryption and decryption. Of course, the sole GPU entropy pool may also be used by CPU-bound implementation of cryptographic applications; in reverse, CPU-bound RBGs and DRBGs can feed random bits in the proposed architecture for on-GPU RSA computations in mixed computing environments.

Proper RSA key generation impose well-known additional constraints to the prime numbers to be used. These RSA-specific primes are further formalized by e.g. the standards and recommendations published by regulatory organizations. RSA key generation, namely the private primes $p$, $q$, composing the public key $N=pq$ and the private exponent $d$, modular inverse of the public exponent $e$ modulo $(p-1)(q-1)$, is completely performed on the GPU with only $e$--when not chosen by the user--and $N$ being transferred back to the CPU host when needed. The depth of the GPU entropy pool and the high-throughput of the MR-TRNG random bits generator make for a very efficient implementation of RSA key generation.

The API provides two GPU kernels respectively for encryption and decryption of messages, applying modular exponentiation. Messages in binary format are split in blocks the length of which matches the number of threads on the GPU used in the proposed architecture. Considering, for instance, 3072-bit long RSA keys, messages are thus split into 3072-bit blocks, each one encrypted and decrypted in parallel by 128 threads within each block. Provided GPUs can process several blocks in parallel several hundreds, even thousands, of message blocks can be efficiently encrypted and decrypted in parallel.
\section{Evaluation}
\label{sec:Evaluation}
 While the architecture proposed in this paper is now provided as a commercial product, the evaluation setup reported here reflects the basic implementation in the original, R\&D experimental work. Namely, we only report on the OpenCL implementation of the MR-TRNG, MR-MOD and MR-RSA kernels.  Several hardware devices deliberately chosen in the performance mid-range were targeted: (i) GPU and (ii) multi-core CPU. Table 1 describes the details of the target hardware.
\begin{table}[h]
\caption{Experimental hardware setup summary.}
\label{tab:example1} \centering
\begin{tabular}{|c|c|}
  \hline
  Implementation & Technical Specs. \\
  \hline
  Multicore CPU & i5 2450M \\
  & 2 cores 4 threads 2.5 GHz \\
  \hline
  GPU & AMD Radeon HD 7670M \\
  & (middle-class GPU for laptops) \\
  & 600-MHz Core speed,  900-Mhz Mem. speed \\
  \hline
  GPU & AMD Radeon HD 6870 \\
  & (middle-class GPU) \\
  & 900-MHz Core speed,  1 GB Mem. \\
  \hline
\end{tabular}
\end{table}
\subsection{Related Work}
 Comparison with related art is not straightforward since different GPU platforms are employed, with different architectural characteristics and performance capabilities. In addition, the rapid pace of progress of manufacturers may render experimental results obsolete quite quickly.

In \cite{Jeske2012} a mixed CUDA/PTX assembler implementation on the high-range Nvidia's GeForce GTX 580 is evaluated against a highly optimized MPIR 2.5.0 bignum library on an Intel i7 processor. The evaluation shows $10^5$ to $250$ modular exponentiation per second when the exponent size varies from 512 bits to 4096 bits, a speed-up of 13x to 3x over an OpenMP CPU implementation.

In \cite{Szerwinski2008} different approaches are proposed and compared to compute asymmetric cryptography (RSA and Elliptic Curve) on Nvidia's GeForce 8800 GTS. The authors show 194 to 419 1024-bit modular exponentiation per second, and 28 to 56 2048-bit modular exponentiation per second. 

The PACE library on the Nvidia's GeForce 9800 GX2--with 2 on-board GPUs--is evaluated in \cite{Giorgi2009}. Experimental results show 45 160-bit modular multiplications per millisecond down to 4 384-bit modular multiplications per millisecond.

In \cite{Harrison2009} the authors devote a sizeable portion of their paper to modular multiplication. They implement Montgomery multiplication in radix representation with the pencil-and-paper algorithm. It is configured so that a single exponent encompasses a CUDA warp (32-threads) in order to maximize thread utilization. Messages are split between two warps dealing with modulus $p$ and  $q$ respectively.  Certain algorithmic optimizations, however, such as squaring and modulo multiplication could not be implemented in this parallel implementation.

Also \cite{Bajard2004} contains a comparative study of innovative variant implementations of RNS-based modular arithmetic in view of RSA encryption and decryption.
\begin{figure}[!h]
  \centering
  \includegraphics[height=8cm, width=8cm]{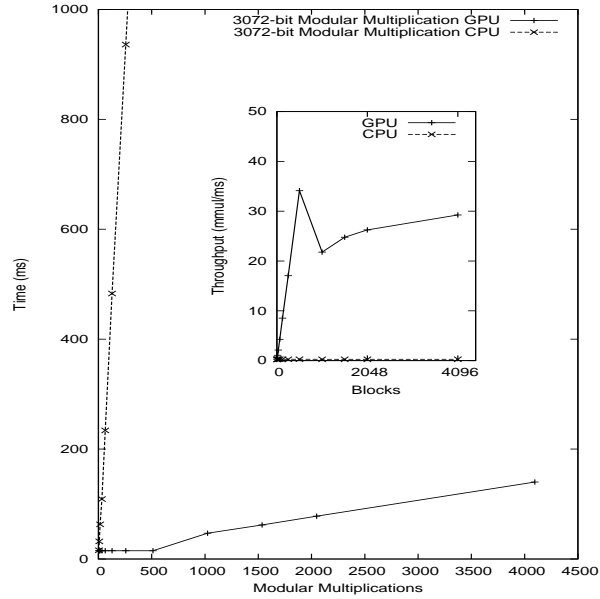}
  \caption{Throughput of the OpenCL implementation for GPU and CPU devices. Execution time for modular multiplication of 3072-bit integers over 128 32-bit RNS channels. Inset: throughput in modular multiplications per millisecond according to number of OpenCL blocks; peak is reached on GPU for the nominal maximum of blocks handled by the AMD card, i.e. 512.}
  \label{fig:example1}
 \end{figure}
Our experimental OpenCL implementation provides results on mid-range hardware along the lines of this previous body of work on high-range devices. As summarized in Figure 1, the MR-MOD layer in the proposed architecture handles up to 3072-bit integer modular multiplications, at performance level comparable to the high-range GPUs previously reported. Doubling the RNS channels number up to the maximum thread-per-block limit afforded by the selected hardware, extends the same performance level to up to 8192-bit integer modular multiplication with the same kernels.

These results point to the feasibility of a complete RSA cryptography trust architecture based on GPU parallel computation. Our additional results, detailed in the next sections, suggest further that GPU may act both as a deep entropy pool for random bit generation and as a consumer of the later randomness for on-board RSA key generation, message encryption and decryption with excellent performance at a relatively low price-point.
\subsection{Uniformly Distributed Random Numbers Generation}
 TestU01 \cite{TestU01} is generally recognized as providing the most complete battery of statistical tests for random number generators.  We submitted MR-TRNG to the so-called ``Big Crush'' battery, a suite of 106 very stringent statistical tests, in the 1.2.1 version of TestU01. As comparison points we reproduce results of McCullough's review of TestU01 \cite{McCullough} along ours:
\begin{table}[h]
\caption{Big Crush tests applied to several RNGs}
\label{tab:bigcrush} \centering
\begin{tabular}{|c|c|}
\hline
RNG &  Failures (CPU Time) \\
\hline
Mersenne Twister & 0 (15:58:25) \\
\hline
Knuth TAOCP & 0 (14:20:26) \\
\hline
Knuth TAOCP 2002 & 0 (14:29:36) \\
\hline
MR-TRNG & 0 (05:17:44) \\
\hline
\end{tabular}
\end{table}
The test file was generated on Nvidia's GPU GeForce GTX 275. In addition, a self-validating kernel was developed to further qualify MR-TRNG. The kernel streamlines FIPS \cite{FIPS:2001:SRC} basic tests right after the generation on the GPU as a form of simplified health test of the generator. As GPUs provide a very high degree of parallelism, MR-TRNG delivers very high-throughput unpredictable streams, in the range of 32-40 GB/s of seed bits, depending on the GPU board. This high performance level doubles up MR-TRNG usage as a complementary deep entropy pool source for traditional CPU-based DRBGs  used in popular cryptographic libraries such as OpenSSL or PolarSSL.

\subsection{Generation, Encryption and Decryption}
 As public key cryptography depends on very large integer multiplications, special attention has been brought to the orchestration of the modular arithmetic, provided by the MR-MOD layer in the proposed architecture, for RSA operations. RSA key generation leverages a parallel implementation of small primes testing (up to the first 10,000 primes) combined with Miller-Rabin compositeness tests--in numbers of iterations as prescribed by the FIPS documentation. Private keys are generated and stay on GPU during all RSA-related calculations.

This flexibility allows for several approaches to parallelizing RSA encryption and decryption of messages. In the experimental evaluation results presented here, the design choice is to break the original message into 3072-bit binary blocks and feed each one of them to an OpenCL block running all the required RNS channel threads for modular exponentiation in parallel. All blocks are submitted for GPU-based RSA encryption and decryption in one single pass. Figure 2 summarizes the results.
\begin{figure}[!t]
  \vspace{-0.2cm}
  \centering
  \includegraphics[ height=8cm, width=8cm]{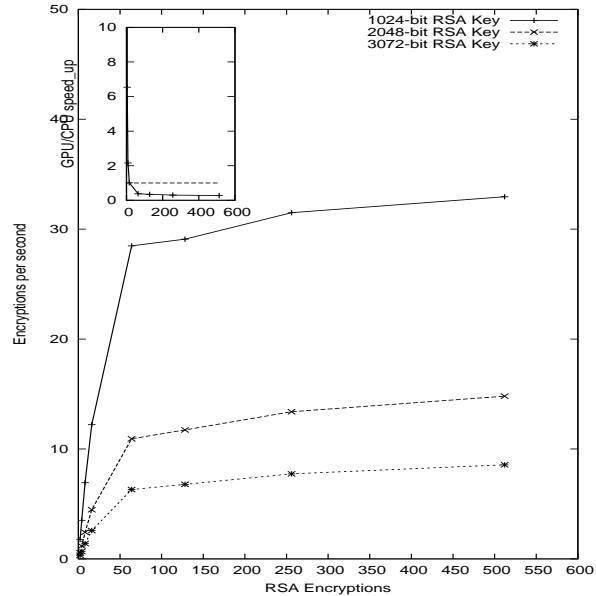}
  \caption{Throughput of the OpenCL implementation for RSA encryptions on GPU. Encryptions of 3072-bit messages, over 128 32-bit RNS channels, for various key lengths. Inset: 0.01x to 16x speed-up improvements over CPU-bound openssl/bn optimized implementation for 3072-bit message and 3072-bit RSA exponent.}
  \label{fig:example2}
  \vspace{-0.1cm}
\end{figure}
\section{Conclusions}
\label{sec:Conclusions}
Experimental results of basic implementations of the proposed GPU-bound MR randomness and cryptography functionality suggest that it is suitable for the dissemination of trust architectures on low-cost, readily available devices. Autonomy at low cost for massively distributed devices, such as routers, laptops, smartphones and tablets, in the required cryptographic key generation and applications may provide a partial response to the challenge brought forward by previously mentioned systemic studies\cite{Lenstra2012,weakkeys2012}. Rather than relying on keys built in at distribution time or delivered in bulk by remote servers, such personal devices would generate high-quality cryptographic keys on demand, leveraging the on-board GPU physical randomness pool. 

The MR Architecture also significantly lowers the cost of key generation and encryption/decryption on personal devices. This in turn may expand usage of cryptographic systems to personal communications, or to many-to-many group communications in dynamic networks.

Finally, further research work towards assessing the physical characterization of the MR-TRNG non-deterministic random bits generator should conclude on the appropriateness of GPU hardware for the design and deployment of Physically Unclonable Function (PUF) systems, a promising solution to additional security issues.
\section*{Acknowledgements}
The authors wish to thank Pr. Jean-Jacques Quisquater, Université catholique de Louvain, for sharing numerous insights and recommendations during the development of this research.

%
%
\bibliographystyle{plain}
\bibliography{massiverand}
\clearpage
\end{document}